\documentclass[journal]{IEEEtran}
\ifCLASSINFOpdf
\else
\fi
\usepackage{color}
\usepackage{epsfig}
\usepackage{amssymb}
\usepackage{amsmath}
\usepackage{amsfonts}
\usepackage[normalem]{ulem}

\usepackage{mwe}
\usepackage{graphbox}

\begin{document}
%
% paper title
% Titles are generally capitalized except for words such as a, an, and, as,
% at, but, by, for, in, nor, of, on, or, the, to and up, which are usually
% not capitalized unless they are the first or last word of the title.
% Linebreaks \\ can be used within to get better formatting as desired.
% Do not put math or special symbols in the title.
\title{Total Variation Bayesian Learning via Synthesis}
%
%
% author names and IEEE memberships
% note positions of commas and nonbreaking spaces ( ~ ) LaTeX will not break
% a structure at a ~ so this keeps an author's name from being broken across
% two lines.
% use \thanks{} to gain access to the first footnote area
% a separate \thanks must be used for each paragraph as LaTeX2e's \thanks
% was not built to handle multiple paragraphs
%

\author{Victor~Churchill,
        Anne~Gelb%,~\IEEEmembership{Fellow,~OSA,}
        %and~Jane~Doe,~\IEEEmembership{Life~Fellow,~IEEE}% <-this % stops a space
\thanks{V. Churchill and A. Gelb are with the Department
of Mathematics, Dartmouth College, Hanover,
NH, 03755 USA e-mail: Victor.A.Churchill.GR@dartmouth.edu.}
\thanks{This work is supported in part by the grants NSF-DMS 1502640, NSF-DMS 1732434, and AFOSR FA9550-18-1-0316.}% <-this % stops a space
%\thanks{J. Doe and J. Doe are with Anonymous University.}% <-this % stops a space
%\thanks{Manuscript received April 19, 2005; revised August 26, 2015.}
}

\maketitle

\begin{abstract}
This paper presents a sparse Bayesian learning algorithm for inverse problems in signal and image processing with a total variation (TV) sparsity prior. Because of the prior used, and the fact that the prior parameters are estimated directly from the data, sparse Bayesian learning often produces more accurate results than the typical maximum \emph{a posteriori} Bayesian estimates for sparse signal recovery. It also provides a full posterior distribution. However, sparse Bayesian learning is only available to problems with a direct sparsity prior or those formed via synthesis.  This paper demonstrates how a problem with a TV sparsity prior can be formulated in a synthesis approach. We then develop a method that combines this synthesis-based TV with the sparse Bayesian learning algorithm and provide numerical examples to demonstrate how our new technique is effectively employed.
\end{abstract}

% Note that keywords are not normally used for peerreview papers.

% For peer review papers, you can put extra information on the cover
% page as needed:
% \ifCLASSOPTIONpeerreview
% \begin{center} \bfseries EDICS Category: 3-BBND \end{center}
% \fi
%
% For peerreview papers, this IEEEtran command inserts a page break and
% creates the second title. It will be ignored for other modes.
\IEEEpeerreviewmaketitle

\section{Introduction}\label{sec:intro}
\IEEEPARstart{I}{n} the synthesis-based approach typically associated with compressed sensing, a set of measurements, $\mathbf{b}$, is collected as a linear combination of an $N$-dimensional signal of interest, $\mathbf{x}$, yielding $\mathbf{b}=\mathbf{A}\mathbf{x}+\mathbf{n}$, where $\mathbf{A}$ is some forward measurement model and $\mathbf{n}$ is a noise vector. Assume $\mathbf{x}$ has a sparse representation in a basis $\mathbf{V}\in\mathbb{R}^{N\times M}$. That is, $\mathbf{x} = \mathbf{V}\mathbf{s}$ with $\mathbf{s}$ sparse, meaning many of its $M$ entries  are zero. While the inversion of $\mathbf{AVs}=\mathbf{b}$ is typically ill-posed, if certain conditions are met, then with high probability $\mathbf{s}$ can be accurately reconstructed from many fewer than $N$ measurements by using $\ell_1$ regularization, \cite{candes2006robust}. The signal of interest $\mathbf{x}$ can then be synthesized as $\mathbf{x}=\mathbf{V}\mathbf{s}$.  Hence this procedure is called the {\em synthesis} approach and $\mathbf{V}$ the synthesis operator, \cite{elad2007analysis}. In this paper, rather than tailoring an operator to achieve an optimally sparse representation, we begin by modifying the simple but rank deficient total variation (TV) operator to conform to the synthesis approach. This enables us to then formulate a sparse Bayesian learning (SBL), \cite{tipping2001sparse}, algorithm for inverse problems in signal and image processing with a TV sparsity prior. We demonstrate through numerical examples that this method outperforms its maximum \emph{a posteriori} counterpart, as has been shown before in \cite{giri2016type,ji2008bayesian,tipping2001sparse,wipf2004sparse} for inverse problems with a direct sparsity prior.

Section \ref{sec:synthesis} explains how to formulate a synthesis approach for TV. In Section \ref{sec:estimation}, we demonstrate how synthetic TV can be used in conjunction with SBL. Section \ref{sec:denoising} provides numerical examples for the archetypal problem of denoising. Concluding remarks and ideas for future research are provided in Section \ref{sec:conclusion}.

\section{Total Variation Regularization via Synthesis}
\label{sec:synthesis}
In part because of its edge-preserving properties, TV regularization, \cite{rudin1992nonlinear}, is a common technique in signal and image processing. TV regularization works by penalizing differences in the value of a signal or an image at adjacent points. It employs the TV operator, $\mathbf{D}\in\mathbb{R}^{(N-1)\times N}$ where
\begin{align}\label{eq:TVtransform}
\mathbf{D}(i,j) = \left\{\begin{array}{cc}
1 & j=i+1\\
-1 & j=i\\
0 & \text{else}\end{array}\right.,
\end{align}
is a scaled finite difference approximation to the gradient.

For piecewise constant signals with just a few jump discontinuities, $\mathbf{D}\mathbf{x}=\mathbf{s}$ is sparse, as in this case $\mathbf{D}$ is an exact transformation to the edge domain. One reason TV regularization is so popular is that in many problems, the signal or image of interest inherently has mostly smooth (and small) variation with just a few edges. Regularization using an approximation to the gradient-like TV has been shown to be a useful tool for restoration  for non-piecewise constant signals as well, even though it is not completely sparsifying in that case. In this paper, we are interested in signals and images with a sparse TV domain, or equivalently having a TV sparsity prior.

Typically in this case, a signal estimate is determined as the minimizer of the $\ell_1$-regularized least squares cost function
\begin{align}\label{eq:analysis}
\arg\min_{\mathbf{x}} \frac12||\mathbf{A}\mathbf{x}-\mathbf{b}||_2^2+\lambda||\mathbf{D}\mathbf{x}||_1.
\end{align}
The first term in (\ref{eq:analysis}) is often called the fidelity term and is minimized when the solution aligns most closely with the given data.  The second term is the imposed sparsity constraint in the TV domain. The regularization parameter $\lambda>0$  balances the fidelity term, the TV sparsity constraint, as well as noise reduction. Unlike the aforementioned synthesis approach, which synthesizes the signal from a sparsity domain estimate, the signal is directly estimated from \eqref{eq:analysis}.  Hence this is called the {\em analysis} approach and $\mathbf{D}$ the analysis operator, \cite{elad2007analysis}. It is important to note that choosing an optimal regularization parameter is generally difficult as there is typically no ground truth to compare with in order to re-tune the parameter. Indeed, chief among the complaints by practitioners in using $\ell_1$ regularization is the difficulty in choosing appropriate regularization parameters, especially in low signal-to-noise ratio (SNR) environments.

Hence in this investigation rather than $\ell_1$ regularization we will employ SBL, which estimates all required parameters from the given data, \cite{tipping2001sparse}. However, SBL is only available to problems formed via synthesis, so we must first find a corresponding synthesis operator for $\mathbf{D}$. If this is achieved, then similar to the process described in the introduction, SBL will recover the sparse TV domain of the signal, and the signal itself will be recovered by applying the TV synthesis operator.

\subsection{Restoring missing information}\label{sec:missing}
A problem quickly arises in developing a synthesis approach to TV, however. Since $\mathbf{D}$ is not invertible (with fewer rows then columns), we do not have the required synthesis operator $\mathbf{V}$ such that $\mathbf{x} = \mathbf{V}\mathbf{s}$. To examine this issue, we cautiously adopt the right pseudoinverse $\mathbf{D}^\dagger=\mathbf{D}^T(\mathbf{D}\mathbf{D}^T)^{-1}$ as the synthesis operator $\mathbf{V}$. Since $\mathbf{D}$ is under-determined and rank deficient, in general $\mathbf{D}^\dagger\mathbf{D}\mathbf{x}\neq\mathbf{x}$. Abstractly, mapping $\mathbf{x}$ to the TV domain and ``back'' via the pseudoinverse may not return to the same space. Therefore an adjustment is required. This adjustment is acknowledged in \cite{chen2001atomic} and \cite{karahanoglu2011signal}, where the authors note that a dictionary of \textit{shifted} heaviside step-functions should act as a TV synthesis operator. In \cite{candes2006robust} (particularly Section I.E), the authors also hint at this phenomenon by introducing a constraint requiring the zeroth Fourier coefficient of the signal to be zero. Corollary 1.4 of \cite{candes2006robust} provides a similar clue, stating that the powerful compressed sensing results presented apply only after the signal has been shifted so its sum is equal to its zeroth Fourier coefficient.

\begin{figure}[t!]
\centering
\includegraphics[width=.5\textwidth]{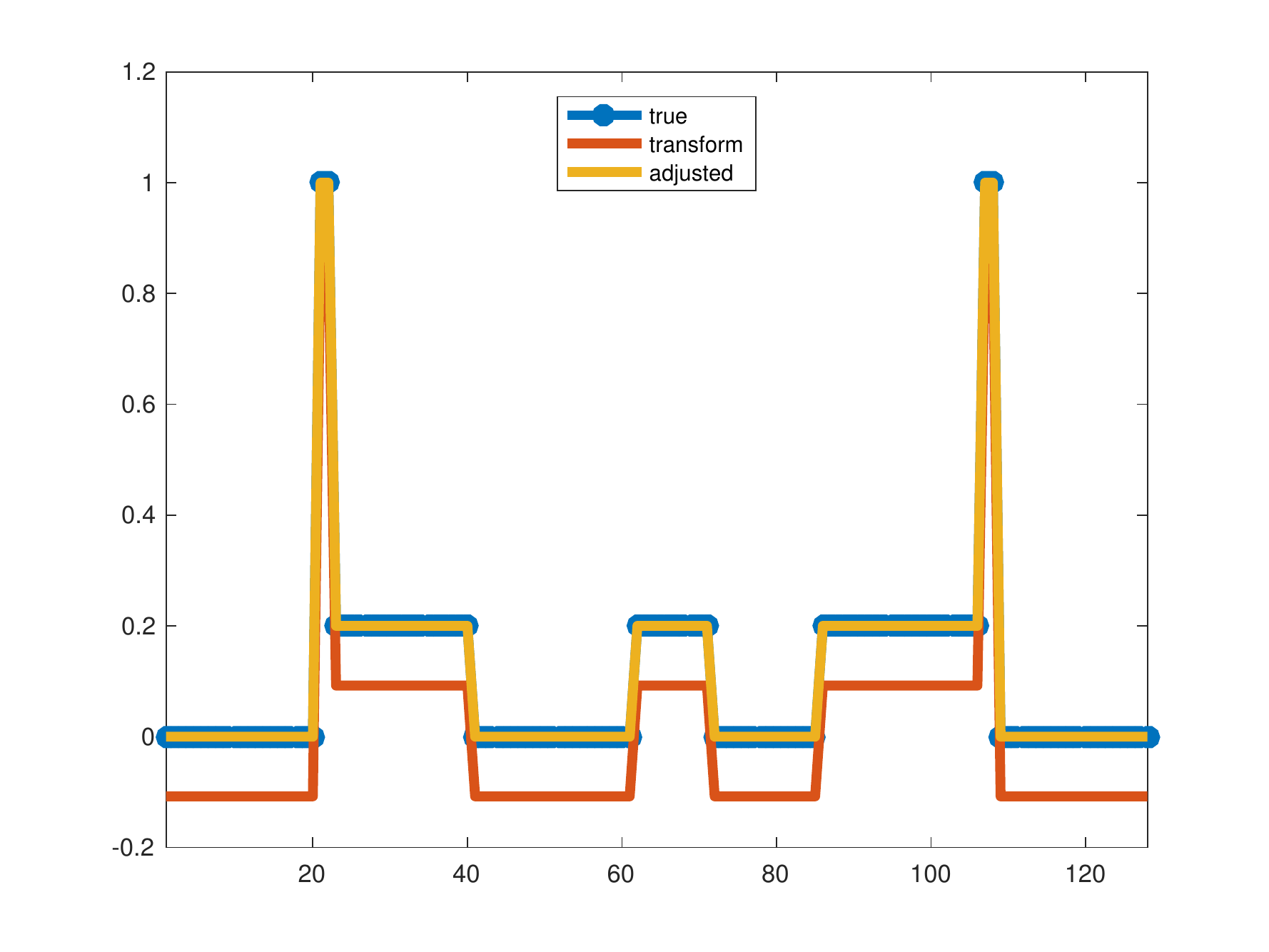}
\caption{One-dimensional slice of the Shepp-Logan phantom before and after adjustment by the mean as in (\ref{eq:adj}).}
\label{fig:edge1D}
\end{figure}

More explicitly, the transformed signal must be adjusted by the mean of the true signal, or equivalently its normalized zeroth Fourier coefficient. A full clarification is immediate from the following simple proof that for \emph{any} vector $\mathbf{x}\in\mathbb{R}^N$,
\begin{align}\label{eq:adj}
\mathbf{x} = \mathbf{D}^\dagger\mathbf{D}\mathbf{x} + \mathbf{\bar{x}},
\end{align}
where $\mathbf{\bar{x}}=\frac1N\sum_{i=1}^N \mathbf{x}_i$. Observe that
\begin{align}\label{eq:pseudoD}
\mathbf{D}^\dagger\mathbf{D}(i,j) = \left\{\begin{array}{cc}
\frac{N-1}{N} & i=j\\
-\frac{1}{N} & \text{else}\end{array}\right..
\end{align}
Then for each $j$,
\begin{align*}
\mathbf{x}_j &= \frac{N-1}{N}\mathbf{x}_j-\frac1N\sum_{\substack{i=1 \\ i\neq j}}^N\mathbf{x}_i+\frac1N\sum_{i=1}^N\mathbf{x}_i = \left(\mathbf{D}^\dagger\mathbf{D}\mathbf{x}\right)_j+\mathbf{\bar{x}}.
\end{align*}

Figure \ref{fig:edge1D} compares $\mathbf{D}^\dagger\mathbf{D}\mathbf{x}$  and (\ref{eq:adj}) for a piecewise constant signal. The decomposition above is equivalent to $\left(\mathbf{D}^\dagger\mathbf{D}+\mathbf{E}\right)\mathbf{x}$ where $\mathbf{E}=\mathbf{I}-\mathbf{D}^\dagger\mathbf{D}$ and $\mathbf{I}$ is the identity matrix. This formulation also allows us to compute $\mathbf{E}$ for \emph{high-order} (HOTV) transforms by replacing $\mathbf{D}$ with an HOTV operator, which will be discussed in future work.

This decomposition can also be applied to images. First define a two-dimensional TV operator $\mathbf{D}_2$. Let $\mathbf{X}$ be an $N\times N$ image. In the anisotropic\footnote{Isotropic TV is not considered here as it has no matrix representation.} TV formulation, edges in the vertical and horizontal directions are separately penalized using
$\mathbf{D}\mathbf{X}$ and $\mathbf{X}\mathbf{D}^T$ with $\mathbf{D}$ as in (\ref{eq:TVtransform}). This can be converted into a single penalty on $\text{vec}(\mathbf{X})$ where $\text{vec}$ is vertical concatenation of the columns of a matrix. Hence $\mathbf{D}_2\in\mathbb{R}^{2N(N-1)\times N^2}$ is defined by
$$\mathbf{D}_2\text{vec}(\mathbf{X}) = \begin{bmatrix}
\mathbf{I}\otimes\mathbf{D}\\
\mathbf{D}\otimes\mathbf{I}
\end{bmatrix}\text{vec}(\mathbf{X})= \begin{bmatrix}\text{vec}(\mathbf{D}\mathbf{X}\mathbf{I})\\\text{vec}(\mathbf{I}\mathbf{X}\mathbf{D}^T)\end{bmatrix},$$
with $\otimes$ the Kronecker product, $\mathbf{I}$ the identity matrix, and the second equality due to Roth's column lemma \cite{roth1934direct}.

\begin{figure}[t!]
\centering
\includegraphics[width=.24\textwidth]{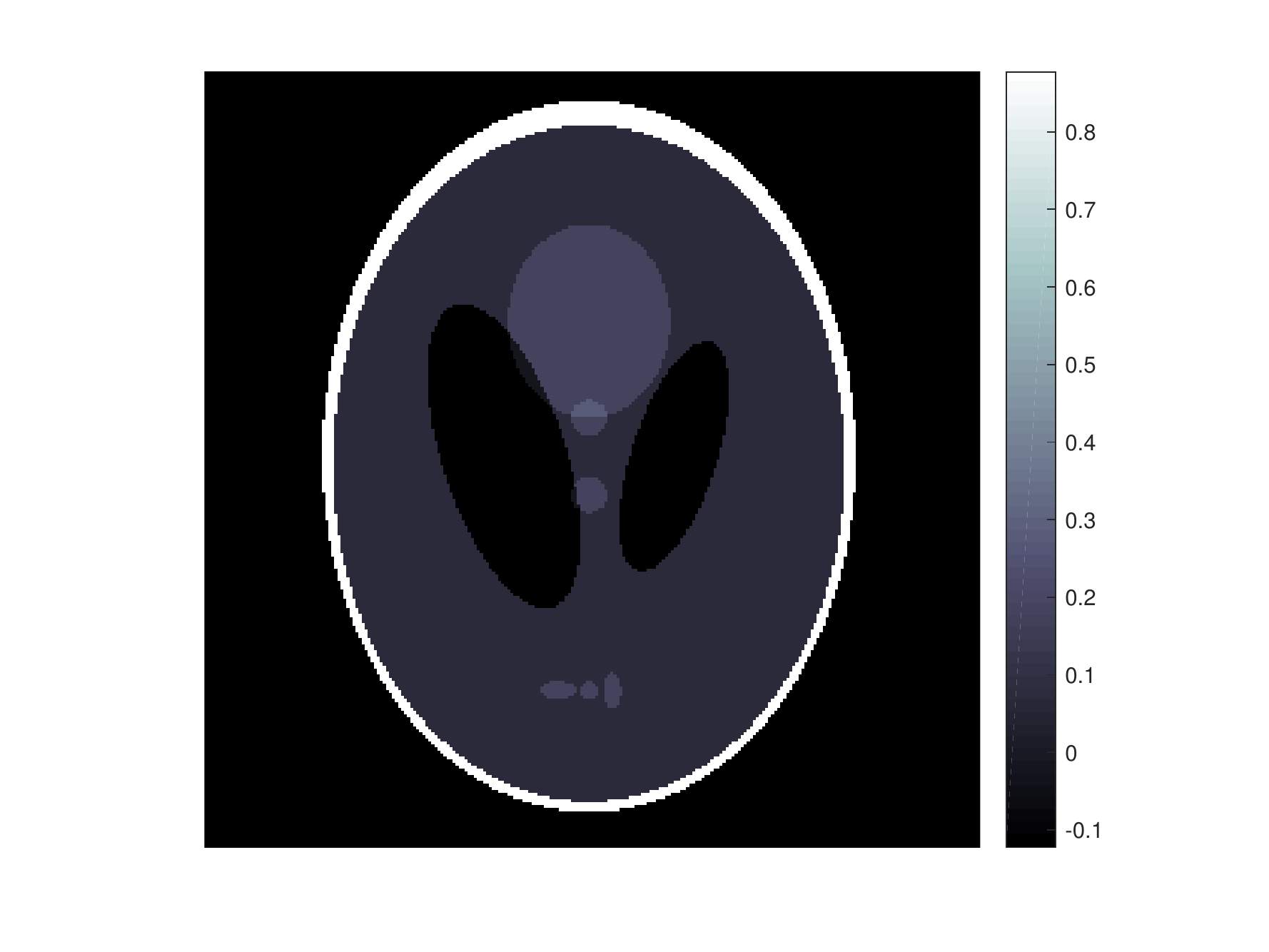}
\includegraphics[width=.24\textwidth]{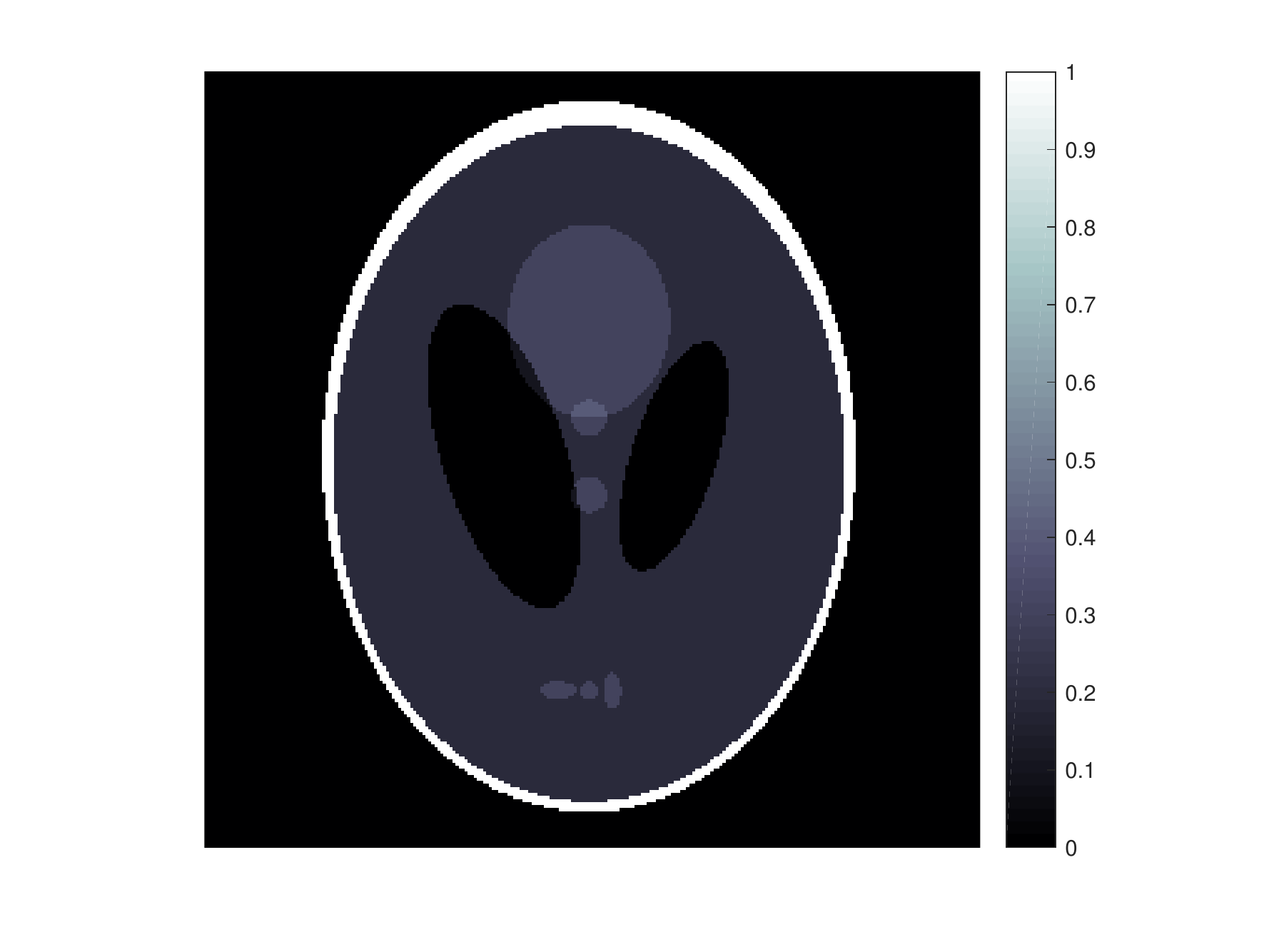}
\caption{Shepp-Logan phantom (left) before and (right) after adjustment by the mean as in (\ref{eq:adj2d}). Note the colorbar.}
\label{fig:edge2D}
\end{figure}

We now prove that, analogous to (\ref{eq:adj}), for \emph{any} image $\mathbf{X}$,
\begin{align}\label{eq:adj2d}
\text{vec}(\mathbf{X}) = \mathbf{D}_2^\dagger\mathbf{D}_2\text{vec}(\mathbf{X})+\mathbf{\bar{X}},
\end{align}
where $\mathbf{\bar{X}} = \frac{1}{N^2} \sum_{i=1}^{N^2}\text{vec}(\mathbf{X})_i$. Observe that
\begin{align}\label{eq:pseudoD2}
\mathbf{D}_2^\dagger\mathbf{D}_2(i,j) = \left\{\begin{array}{cc}
\frac{N^2-N}{N^2} & i=j\\
-\frac{1}{N^2} & \text{else}\end{array}\right..
\end{align}
Then for each $j$
\small
\begin{align*}
\text{vec}(\mathbf{X})_j &= 
\frac{N^2-1}{N^2}\text{vec}(\mathbf{X})_j
- \frac{1}{N^2}\sum_{\substack{i=1\\i\neq j}}^{N^2}\text{vec}(\mathbf{X})_i +\frac{1}{N^2}\sum_{i=1}^{N^2}\text{vec}(\mathbf{X})_i \\ &= \left(\mathbf{D}_2^\dagger\mathbf{D}_2\text{vec}(\mathbf{X})\right)_j + \mathbf{\bar{X}}.
\end{align*}
\normalsize Figure \ref{fig:edge2D} compares $\mathbf{D}_2^\dagger\mathbf{D}_2\text{vec}(\mathbf{X})$ and (\ref{eq:adj2d}) for the Shepp Logan phantom, \cite{shepp1974fourier}.

The above information does not provide much insight on how to obtain the mean of the true signal required in \eqref{eq:adj} and \eqref{eq:adj2d}. However, there are a few examples where this constant can be recovered. One case is when the zeroth Fourier coefficient is $\mathbf{\hat{x}}_0=0$, as was mentioned in \cite{candes2006robust}. This occurs, e.g., when $\mathbf{x}$ is the gradient of another signal, \cite{chartrand2017numerical}. Another case is when Fourier data is collected, hence we may simply add back in the normalized zeroth coefficient. Many imaging systems collect Fourier data, e.g. magnetic resonance imaging and synthetic aperture radar. Finally, even in cases where $\mathbf{\bar{x}}$ is not explicitly known, it may still be possible to make a reasonable approximation. Under the transformation $\mathbf{D}^\dagger\mathbf{D}$, signal shape is preserved with only a constant missing. Therefore, subtracting the transform value $\mathbf{D}^\dagger\mathbf{D}\mathbf{x}$ at a zero intensity location in the true signal from every pixel correctly adjusts the signal. This approach can be applied regardless of the forward model. The assumption that an area of zero intensity is known is reasonable in medical imaging applications, where a large buffer of zero intensity is commonly placed around the object being imaged. The issue of noise in the shift will be addressed in Section \ref{sec:denoising}. Using the above information, we can now formulate a synthesis-based Bayesian approach to TV regularization.

\section{Bayesian Estimation}\label{sec:estimation}
In synthesis, we are concerned with the problem of recovering a sparse signal $\mathbf{s}$ from noisy measurements
\begin{align}\label{eq:model}
\mathbf{b} &= \mathbf{AD}^\dagger\mathbf{s}+\mathbf{n}.
\end{align}
We will assume $\mathbf{n}$ is zero-mean Gaussian noise with unknown variance $\nu^2$. We model in this way with the intent to recover $\mathbf{s}=\mathbf{Dx}$ and synthesize $\mathbf{x}$ as described in Section \ref{sec:synthesis}. Since $\mathbf{s}$ is sparse for piecewise constant $\mathbf{x}$, this inverse problem is referred to as sparse signal recovery (SSR). Recently there has been interest in Bayesian approaches to improving the accuracy of solutions to SSR, \cite{babacan2008parameter,babacan2010bayesian,giri2016type,ji2008bayesian,tipping2001sparse,wipf2004sparse}. In these approaches, the assumption that $\mathbf{s}$ is sparse is used to inform a prior probability distribution and then using the given data and forward model a posterior distribution for $\mathbf{s}$ is sought. These Bayesian methods can be divided into two categories that encompass many popular SSR methods. In type-I, or maximum \textit{a posteriori} (MAP), Bayesian estimation uses a fixed prior. This category includes the popular $\ell_1$ regularization method, \cite{tibshirani1996regression}. In type-II, or evidence maximization, Bayesian estimation employs a flexible and hierarchical parametrized prior that is learned from the data. An exhaustive empirical comparison of both methods was performed in \cite{giri2016type}, where it was concluded that type-II estimates are typically more accurate than corresponding type-I estimates. In addition to improved accuracy, the type-II framework provides a full posterior distribution as opposed to only a point estimate given by type-I methods. The type-II framework also incorporates data-driven parameter estimation into the algorithm. This is crucial as choosing the regularization parameter in type-I schemes is frequently difficult and problem-dependent without oracle knowledge, requiring user input and investigation. Below we briefly describe both of these approaches to TV regularization via synthesis.

\subsection{MAP estimation (Type I)}\label{sec:MAP}

As in \cite{ji2008bayesian}, assuming the entries of $\mathbf{b}$ in \eqref{eq:model} are independent, we have the Gaussian likelihood model
\begin{align}\label{eq:glm}
p(\mathbf{b}|\mathbf{s},\nu^2) &= (2\pi\nu^2)^{-J/2}\exp\left(-\frac{1}{2\nu^2}||\mathbf{AD}^\dagger\mathbf{s}-\mathbf{b}||_2^2\right),
\end{align}
where $J$ is the length of the measurement vector $\mathbf{b}$. We formulate the assumption that $\mathbf{s}$ is sparse by using a fixed sparsity-encouraging prior, e.g. the Laplace density function
\begin{align}\label{eq:laplaceprior}
p(\mathbf{s}|\mu) &= \left(\frac{\mu}{2}\right)^{N-1}\exp\left(-\mu||\mathbf{s}||_1\right).
\end{align}
Note that $N-1$ is the length of $\mathbf{s}$. There are many sparsity-encouraging priors sometimes referred to as super-Gaussians as they are characterized by fat tails and a sharp peak at zero. Using Bayes' theorem we compute the MAP estimate as
\begin{align}\label{eq:MAP}
\begin{split}
\mathbf{s}_{MAP} &=\arg\max_\mathbf{s} p(\mathbf{s}|\mathbf{b})= \arg\max_\mathbf{s} p(\mathbf{b}|\mathbf{s},\nu^2)p(\mathbf{s}|\mu) \\&= \arg\min_\mathbf{s}\left\{\frac12||\mathbf{AD}^\dagger\mathbf{s}-\mathbf{b}||_2^2+\nu^2\mu||\mathbf{s}||_1\right\}.
\end{split}
\end{align}
The prior knowledge of sparsity parameter $\mu$ and noise parameter $\nu^2$ corresponds to the assertion of a regularization parameter usually called $\lambda=\nu^2\mu$. Therefore, since $\mathbf{D}\mathbf{x} = \mathbf{s}$, from (\ref{eq:adj}) the TV synthesis approach gives the MAP estimate
\small
\begin{align}\label{eq:1DMAP}
\mathbf{x}_{MAP} = \mathbf{D}^\dagger\cdot\arg\min_\mathbf{s} \left\{\frac12||\mathbf{A}\mathbf{D}^\dagger\mathbf{s}-\mathbf{b}||_2^2+\lambda||\mathbf{s}||_1\right\}+\mathbf{\bar{x}}.
\end{align}
\normalsize Equations (\ref{eq:analysis}) and (\ref{eq:1DMAP}) retrieve nearly equal estimates, \cite{elad2007analysis}. For images, replace $\mathbf{s}$ with $\text{vec}(\mathbf{S})$ and $\mathbf{D}$ with $\mathbf{D}_2$.

\subsection{Total variation Bayesian learning (Type II)}\label{sec:typeii}
In type-II Bayesian estimation, instead of a fixed sparsity-inducing prior on $\mathbf{s}$, an empirical prior characterized by flexible parameters that must be estimated from the data is used. In this investigation we focus on one type-II method called sparse Bayesian learning (SBL), \cite{tipping2001sparse}, which was used in Bayesian compressed sensing, \cite{ji2008bayesian}. SBL is only available to problems formed via synthesis or directly sparse problems, hence our derivation of a synthesis approach to TV. The reason we seek to employ SBL is that in many cases it has been shown empirically and theoretically to be superior in terms of accuracy to type-I estimates, \cite{faul2002analysis,giri2016type,wipf2004sparse,wipf2005norm}. Theoretical analysis in \cite{rao2006comparing} and \cite{wipf2005norm}, shows that SBL provides a closer approximation to the $\ell_0$ norm of the sparse signal than the $\ell_1$ norm. For the noiseless case, it was shown in \cite{wipf2004sparse} that the global minimum of the effective SBL cost function is achieved at a solution such that the posterior mean equals the maximally sparse solution. Furthermore, local minima are achieved at sparse solutions, regardless of noise. Empirically, \cite{giri2016type} shows that SBL achieves superior SSR results compared to $\ell_1$, reweighted $\ell_1$, and reweighted $\ell_2$ regularization (see \cite{candes2006robust,candes2008enhancing,chartrand2008iteratively}, respectively). This is further supported by multi-run testing in \cite{ji2008bayesian}. In addition, SBL provides a full posterior distribution versus a point estimate, and automatically estimates all parameters from the given data requiring no user input.

Hence SBL will be used in an attempt to more accurately detect the sparse representation $\mathbf{s}=\mathbf{Dx}$ than a MAP estimate, e.g. (\ref{eq:MAP}).
The signal is then synthesized using an approximation to \eqref{eq:adj} or \eqref{eq:adj2d} as described in Section \ref{sec:synthesis}. The description of SBL below comes from \cite{ji2008bayesian} and \cite{tipping2001sparse}.

First we develop a parametrized prior on $\mathbf{s}$. Because Gaussian noise is assumed in (\ref{eq:model}), we define a zero-mean Gaussian prior on each element of $\mathbf{s}$
\begin{align*}
p(\mathbf{s}|\mathbf{a}) = \prod_{i=1}^{N-1}\mathcal{N}(\mathbf{s}_i|0,\mathbf{a}_i^{-1}),
\end{align*}
where $\mathbf{a}_i$ is the inverse variance. We then define a minimally informative Gamma prior over $\mathbf{a}$ as
\begin{align*}
p(\mathbf{a}|a,b)=\prod_{i=1}^{N-1}\Gamma(\mathbf{a}_i|a,b).
\end{align*}
Finally, we marginalize over the hyperparameters $\mathbf{a}$ to obtain the overall prior on $\mathbf{s}$ as
\begin{align}\label{eq:overallprior}
p(\mathbf{s}|a,b) = \prod_{i=1}^{N-1}\int_0^\infty\mathcal{N}(\mathbf{s}_i|0,\mathbf{a}_i^{-1})\Gamma(\mathbf{a}_i|a,b)d\mathbf{a}_i.
\end{align}
Each integral being multiplied in (\ref{eq:overallprior}) is distributed via the Student's $t$-distribution, which, for suitable $a$, $b$, is strongly peaked at $\mathbf{s}_i=0$. Therefore this prior favors $\mathbf{s}_i$ being zero, hence encouraging sparsity. We also impose a Gamma prior $\Gamma(\beta|c,d)$ on $\beta=\frac{1}{\nu^2}$. Only point estimates are needed for $\mathbf{a}$ and $\beta$, so we simply set $a,b,c,d=0$ implying uniform hyperpriors on a logarithmic scale for $\mathbf{a}$ and $\beta$, \cite{tipping2001sparse}.

Given the prior above, the posterior distribution for $\mathbf{s}$ can be solved for analytically as a multivariate Gaussian distribution
\begin{align*}
p(\mathbf{s}|\mathbf{b},\mathbf{a},\beta) &=\mathcal{N}(\mathbf{s}|\mathbf{m},\Sigma),
\end{align*}
with mean and covariance matrix given by
\begin{align}\label{eq:mean}
\mathbf{m} &= \beta\Sigma(\mathbf{AD^\dagger})^T\mathbf{b},
\end{align}
\begin{align}\label{eq:cov}
\Sigma &= \left(\beta(\mathbf{AD^\dagger})^T\mathbf{AD^\dagger}+\mathbf{\Lambda}\right)^{-1},
\end{align}
where $\mathbf{\Lambda}=\text{diag}(\mathbf{a})$, \cite{bishop2006pattern}. The hyperparameters $\mathbf{a}$ and $\beta$ are now learned from the data. Marginalizing over $\mathbf{s}$, the marginal log-likelihood for $\mathbf{a}$ and $\beta$ is
\begin{align}\label{eq:L}
\begin{split}
\log p(\mathbf{y}|\mathbf{a},\beta) &= \log\int p(\mathbf{y}|\mathbf{g},\beta)p(\mathbf{g}|\mathbf{a})d\mathbf{g}\\&=-\frac12\left(J\log 2\pi + \log |\mathbf{C}| +\mathbf{y}^t\mathbf{C}^{-1}\mathbf{y}\right),
\end{split}
\end{align}
with $\mathbf{C}=\beta^{-1}\mathbf{I}+\mathbf{AD^\dagger} \mathbf{\Lambda}^{-1}(\mathbf{AD^\dagger})^T$, \cite{bishop2006pattern}.  Note that (\ref{eq:L}) cannot be maximized in closed form. In \cite{tipping2001sparse}, a maximum likelihood approximation is employed that uses the point estimates for $\mathbf{a}$ and $\beta$ to maximize (\ref{eq:L}), and is implemented via the {\em expectation-maximization} (EM) algorithm, \cite{dempster1977maximum}. In particular, the update for $\mathbf{a}$ to maximize (\ref{eq:L}) is
\begin{align}\label{eq:a}
\mathbf{a}_i^{\text{(new)}}=\frac{\gamma_i}{\mathbf{m}_i^2}
\end{align}
for each $i$, with $\mathbf{m}_i$ the $i$th posterior mean weight from (\ref{eq:mean}) and $\gamma_i = 1-\mathbf{a}_i\Sigma_{ii}$ with $\Sigma$ from (\ref{eq:cov}). For $\beta$, the update is
\begin{align}\label{eq:beta}
\beta^{\text{(new)}}=\frac{M-\sum_i\gamma_i}{||\mathbf{b}-\mathbf{AD}^\dagger\mathbf{m}||_2^2}.
\end{align}
Appendix A of \cite{tipping2001sparse} gives details on the derivation of these terms. Observe that $\mathbf{a}^{\text{(new)}}$ and $\beta^{\text{(new)}}$ are functions of $\mathbf{m}$ and $\Sigma$, and vise versa. The EM algorithm iterates between (\ref{eq:mean}) and (\ref{eq:cov}), and (\ref{eq:a}) and (\ref{eq:beta}) until a convergence criterion is satisfied. Due to the properties of the EM algorithm, SBL is globally convergent, i.e. each iteration is guaranteed to reduce the cost function, \cite{wipf2004sparse}. It has been observed that most $\mathbf{a}_i\rightarrow\infty$, corresponding to a sparse result with $\mathbf{s}_i\approx0$ for most $i$.

The signal of interest $\mathbf{x}$ is finally recovered as the mean of the multivariate Gaussian posterior distribution given by
\begin{align*}
\mathbf{x}^* &= \mathbf{D}^\dagger\mathbf{m}^* + \mathbf{\bar{x}},
\end{align*}
where $\mathbf{\bar{x}}$ is obtained using a method from Section \ref{sec:missing}. Here $\mathbf{m}^*$ is the final $\mathbf{m}$ from (\ref{eq:mean}) once the convergence criterion is attained. Note that the final $\Sigma$ is the covariance matrix of the posterior density function for $\mathbf{s}$, not $\mathbf{x}$.

While this method has been shown to achieve highly accurate sparse restorations as well as providing the advantage of automatically estimating the parameters of the model and providing a full density, \cite{giri2016type,tipping2001sparse}, for each iteration it requires the inversion of the $(N-1)\times(N-1)$ covariance matrix $\Sigma$, which scales to $\mathcal{O}(N^3)$ operations, which is inefficient for large $N$. Therefore fast algorithms developed in \cite{faul2002analysis,tipping2003fast} are used in the forthcoming two-dimensional numerical experiment. Although these algorithms are based on the same cost function (\ref{eq:L}), we notice an accuracy discrepancy in our empirical testing. Hence future investigations will focus on developing  optimally accurate and fast implementations. Similarly for fast implementation, $\mathbf{D}^\dagger$ should be pre-computed.

\section{Numerical Results}\label{sec:denoising}
\begin{figure}[t!]
\centering
\includegraphics[width=.5\textwidth]{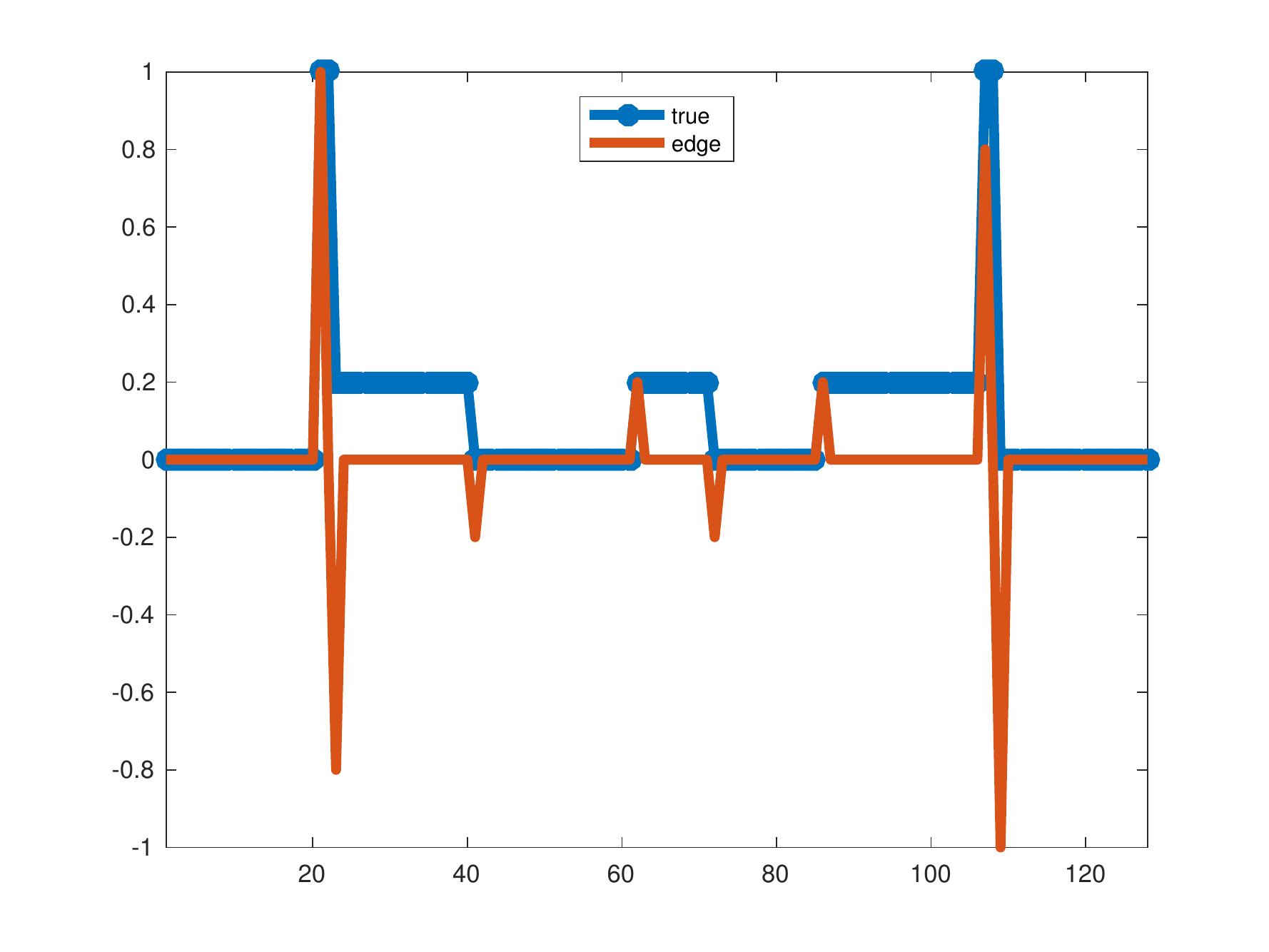}
\caption{Cross-sectional slice of Shepp-Logan phantom with edge.}
\label{fig:D}
\end{figure}

As a proof of concept, we test out this new method on the classical problem of denoising, which epitomizes the difficulty in balancing fidelity, sparsity, and noise reduction. In denoising, $\mathbf{A}$ is the identity, meaning we collect a noisy signal $\mathbf{b} = \mathbf{x}+\mathbf{n}$, and regularize by the TV sparsity of the signal to return a result more faithful to the unknown ground truth image. We compare the resulting reconstructions from (\ref{eq:analysis}) and the proposed TV SBL procedure. Figure \ref{fig:D} shows a horizontal cross-section of Shepp-Logan phantom which we will test on, as well as its edge map (a non-uniform spike train $\mathbf{s}$), which is what the Bayesian learning approach actually recovers. Note that the sparsity level is $8$ edges to $128$ total entries. The noise level in the collected data is measured by signal-to-noise ratio defined
\begin{align}
SNR = 20\cdot\log_{10}\left(\sqrt{\frac{\sum_{i=1}^N \mathbf{x}_i^2}{\sum_{i=1}^N\mathbf{n}_i^2}} \right).
\end{align}
We compare the reconstructions using the relative error defined
\begin{align}
RE(\mathbf{x}^*) = \frac{||\mathbf{x}^* - \mathbf{x}||_2}{||\mathbf{x}||_2},
\end{align}
where $\mathbf{x}^*$ is the restoration and $\mathbf{x}$ is the ground truth. To synthesize and adjust the resulting SBL restoration as described in Section \ref{sec:synthesis}, we add the mean of the noisy signal, which in this case is an unbiased estimator of the ground truth mean. Since we know the ground truth in this case we can optimize the regularization parameter $\lambda$ in (\ref{eq:analysis}) to minimize the relative error. We show this best-case scenario while noting that without oracle knowledge of the signal, this optimal result may be difficult to obtain in real-world examples. Recall that the proposed Bayesian learning approach requires no parameter inputs, only the forward model and data. 

\begin{figure}[t!]
\centering
\includegraphics[width=.5\textwidth,align=c]{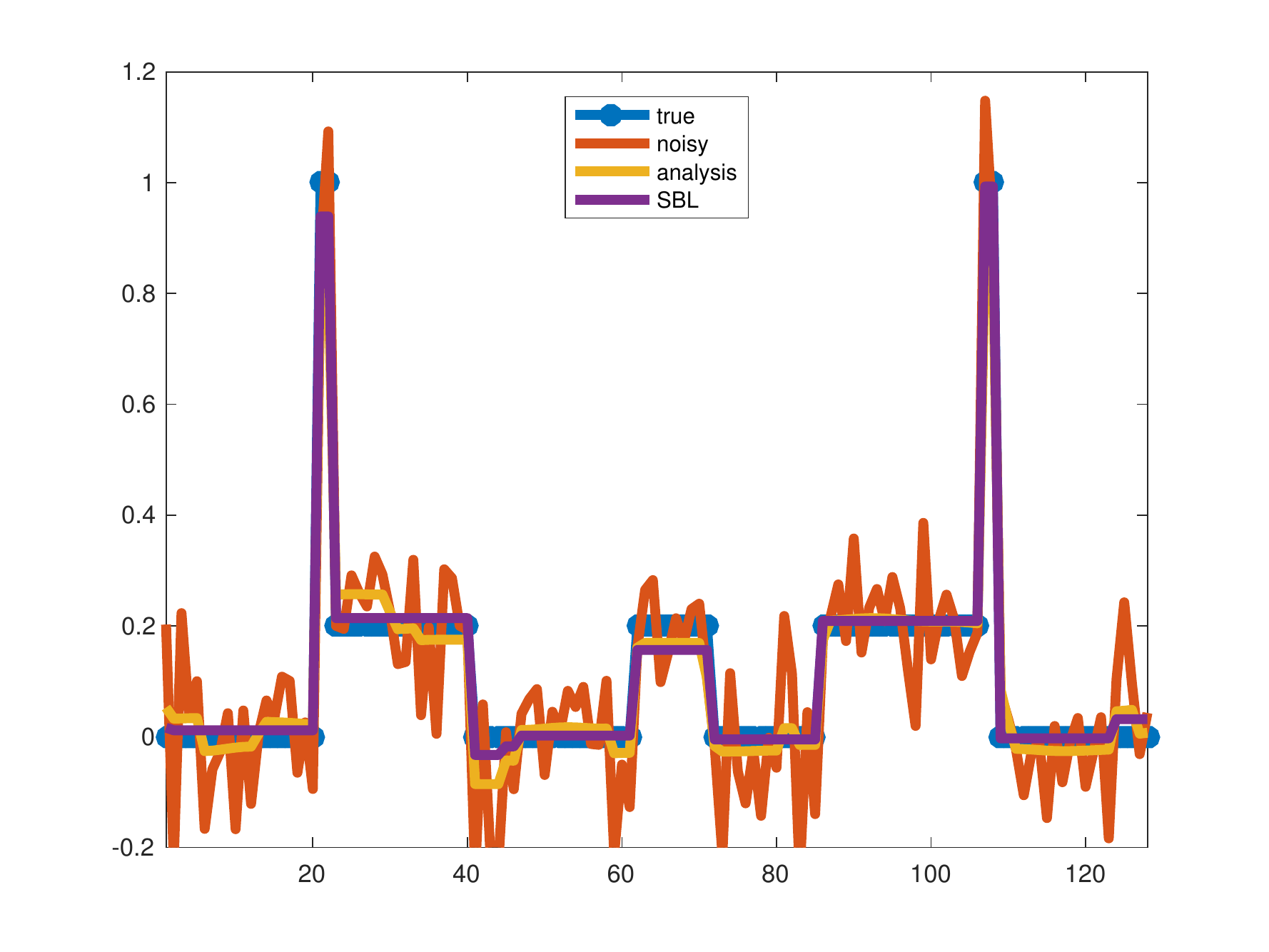}
\includegraphics[width=.5\textwidth,align=c]{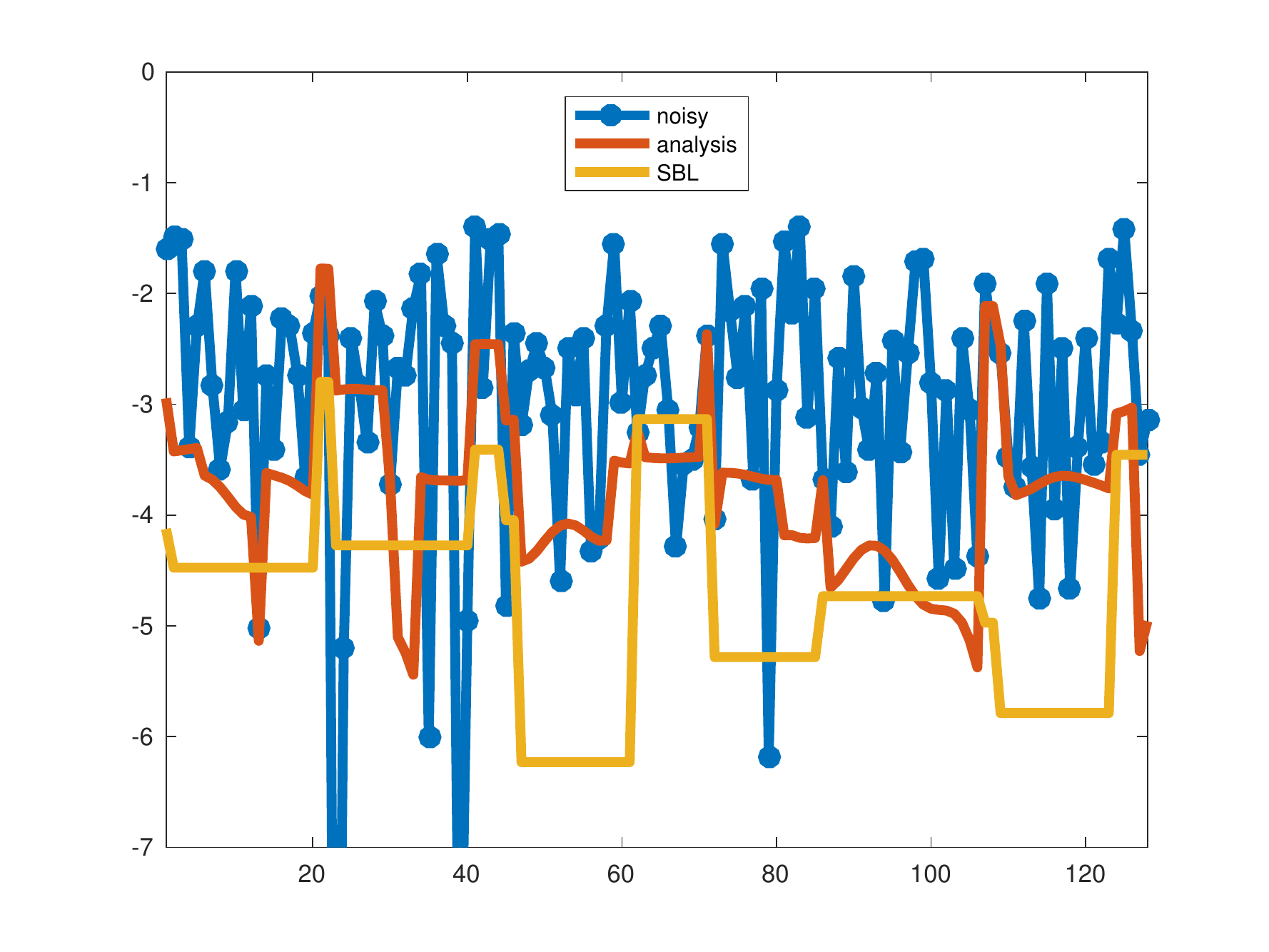}
\caption{TV denoising with SNR$=6.4$. (top) restorations, (bottom) log error. `True' is ground truth, `noisy' is the collected data, `analysis' is the result from (\ref{eq:analysis}), and `SBL' is the result from the proposed algorithm.}
\label{fig:denoising_1D}
\end{figure}

\begin{figure}[t!]
\centering
\includegraphics[width=.5\textwidth,align=c]{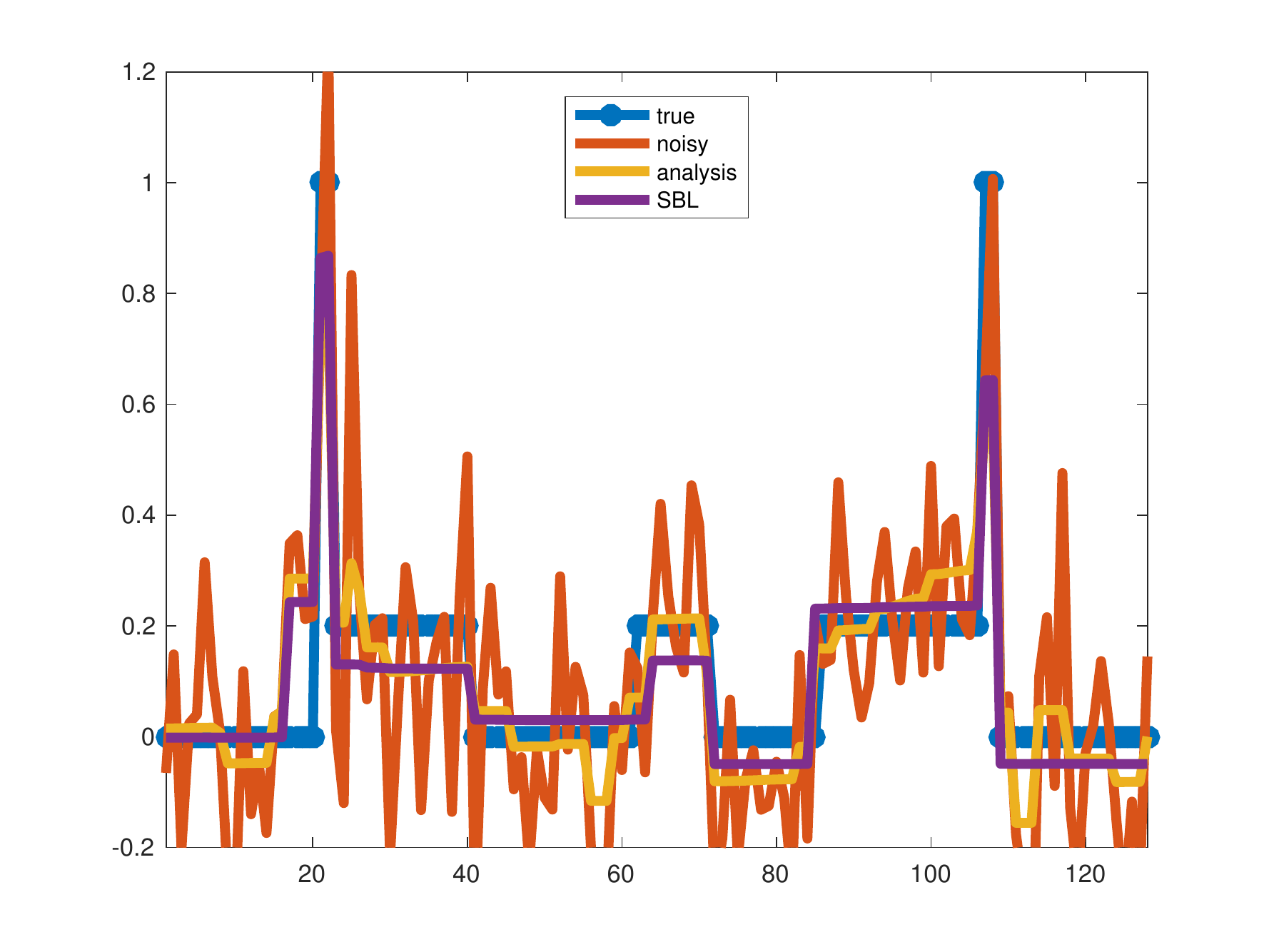}
\includegraphics[width=.5\textwidth,align=c]{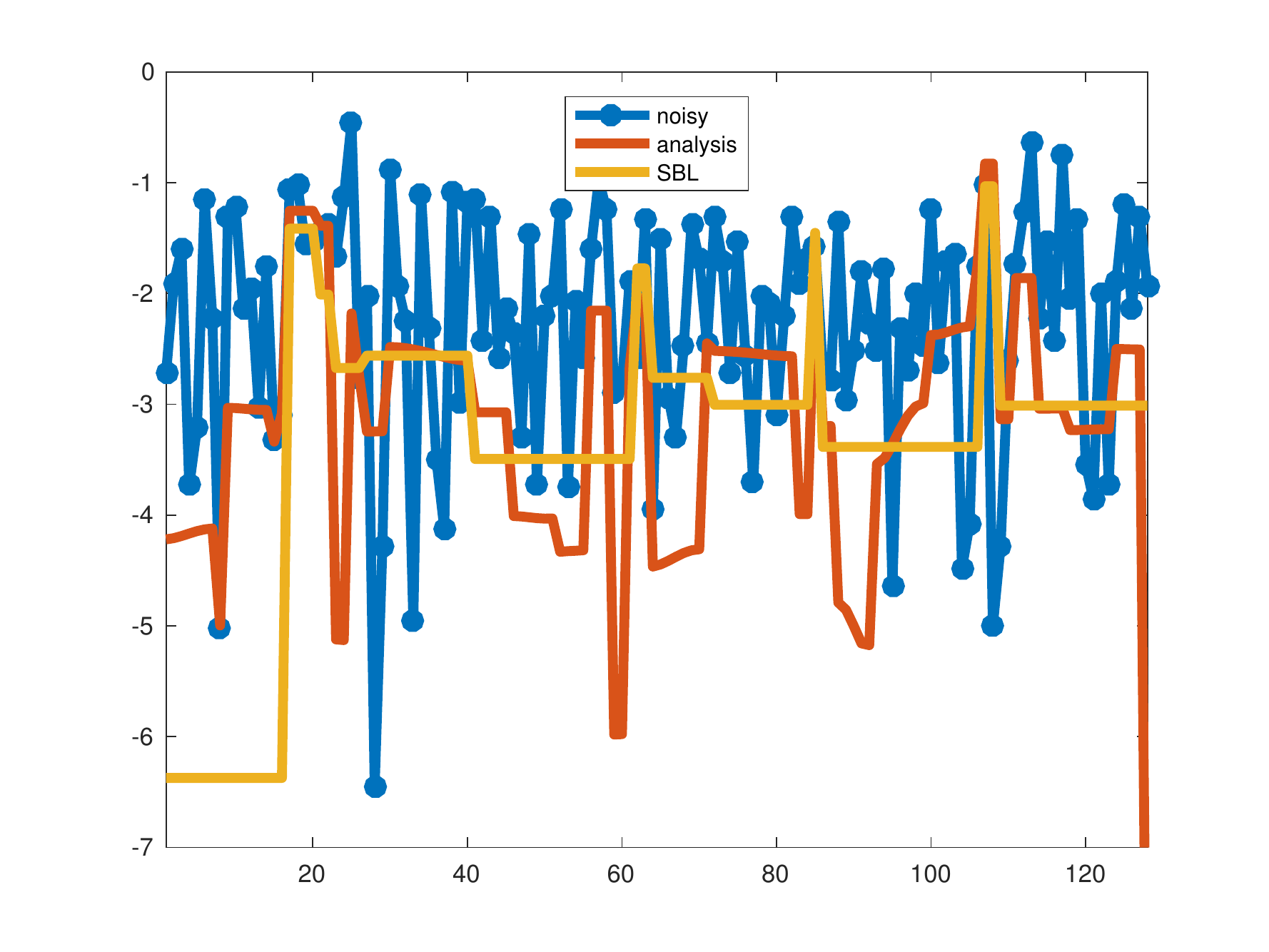}
\caption{TV denoising with SNR$=0.9$. (top) restorations, (bottom) log error. `True' is ground truth, `noisy' is the collected data, `analysis' is the result from (\ref{eq:analysis}), and `SBL' is the result from the proposed algorithm.}
\label{fig:denoising_1D_more}
\end{figure}

\begin{figure}[t!]
\centering
\includegraphics[width=.24\textwidth]{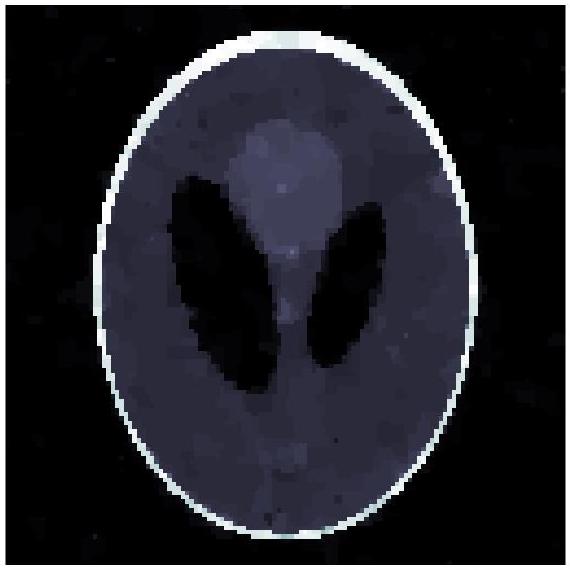}
\includegraphics[width=.24\textwidth]{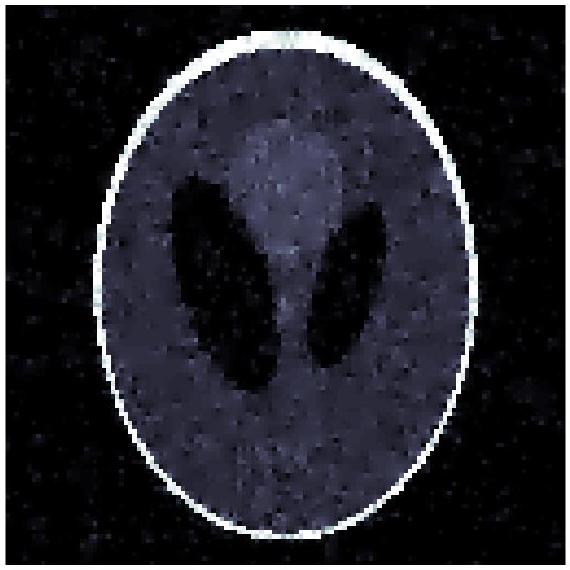}
\caption{TV image denoising example: (left)  Restoration using (\ref{eq:analysis})  with relative-error-minimizing regularization parameter, (right) SBL restoration.}
\label{fig:denoising_2D}
\end{figure}

Figure \ref{fig:denoising_1D} shows a comparison of the reconstructions and log error plots from data with $SNR\approx6.4$. The relative errors were $.1909$ for (\ref{eq:analysis}) and $.0868$ using the SBL approach. We see a significant improvement in accuracy both near edges and in smooth regions. Figure \ref{fig:denoising_1D_more} shows results of this experiment using even more noise. This time, $SNR\approx0.9$, meaning that there is more noise than signal in the collected data. The Bayesian learning approach again outperforms the standard TV approach, achieving relative error of $.3873$ compared with $.4775$. Figure \ref{fig:denoising_2D} shows an image denoising example using the full Shepp-Logan phantom with $SNR\approx8.5$. It is apparent that the Bayesian learning approach does not outperform (\ref{eq:analysis}), which we suspect is due to the fast algorithms employed, \cite{faul2002analysis,tipping2003fast}, and the shortcuts used to increase the speed. This will be investigated in future work.

\section{Conclusion}\label{sec:conclusion}

In this paper we reformulated the classic TV-regularized inverse problem via synthesis by clarifying the missing true signal mean constant when using the right pseudo-inverse as a synthesis operator. This allowed us to explore a sparse Bayesian learning estimation procedure that is only available for synthetically and directly sparse problems. Our results show that these methods show promise because of their accuracy as well as the provision of data-driven parameter estimation. However, they are not yet efficient enough for large problems. Future investigations will include efforts to improve efficiency, perhaps by pre-processing with prior information. We will also develop methods to determine missing shift parameters for other regularization operators, such as HOTV, and how they can be approximated from given data. This will improve accuracy and allow for low resolution environments, thereby increasing efficiency.

%\nocite{*}
\bibliographystyle{acm}
\bibliography{refs}

\end{document}